\documentclass[pre,twocolumn,amsmath,amssymb,floatfix,,superscriptaddress]{revtex4-1}

\usepackage{graphicx}
\usepackage{dcolumn}
\usepackage{bm}

\usepackage{lipsum}

\usepackage[FIGTOPCAP,tight,raggedright,nooneline]{subfigure}
\usepackage{float}
\graphicspath{{images/}{plots/}}

\renewcommand{\vec}[1]{\bm{#1}}

\newcommand{\avg}[1]{\ensuremath{\left< #1 \right>}}

\newcommand{\abs}[1]{\ensuremath{\left\vert#1\right\vert}}

\begin{document}

    \title{How many longest increasing subsequences are there?}
    \author{Phil Krabbe}
    \email{phil.krabbe@uol.de}
    \affiliation{Institut f\"ur Physik, Universit\"at Oldenburg, 26111 Oldenburg, Germany}
    \author{Hendrik Schawe}
    \email{hendrik.schawe@cyu.fr}
    \affiliation{Laboratoire de Physique Th\'{e}orique et Mod\'{e}lisation, UMR-8089 CNRS, CY Cergy Paris Universit\'{e}, 95000 Cergy, France}
    \affiliation{Institut f\"ur Physik, Universit\"at Oldenburg, 26111 Oldenburg, Germany}
    \author{Alexander K. Hartmann}
    \email{a.hartmann@uol.de}
    \affiliation{Institut f\"ur Physik, Universit\"at Oldenburg, 26111 Oldenburg, Germany}
    \date{\today}

    \begin{abstract}
        We study the entropy $S$ of longest increasing subsequences (LIS), i.e.,
        the logarithm of the number of distinct LIS. We consider two ensembles
        of sequences, namely random permutations of integers and
        sequences drawn i.i.d.\ from a limited number of
        distinct integers. Using sophisticated algorithms, we are able to
        exactly count the number of LIS for each given sequence.
        Furthermore, we are not only measuring averages and variances for the
        considered ensembles of sequences, but we sample
        very large parts of the probability
        distribution $p(S)$ with very high precision.
        Especially, we are able to observe the tails of extremely
        rare events which occur with probabilities smaller than $10^{-600}$.
        We show that the distribution of the entropy of the LIS is approximately
        Gaussian with deviations in the far tails, which might vanish
        in the limit of long sequences. Further we propose a
        large-deviation rate function which fits best to our observed data.
    \end{abstract}

    \maketitle

    \section{Introduction}
        Imagine a game of numbers: Given a sequence of $n$ numbers, mark the largest
        subset of numbers
        such that every marked number is larger (or equal) to all marked
        numbers appearing left of it in the sequence. The marked numbers will
        be a \emph{(weakly) increasing subsequence}. The number of marked elements is called the \emph{length} $l$.
        If the subsequence maximizes $l$ over all possible subsequences  it is called a
        \emph{longest (weakly) increasing subsequence} (LIS) \cite{romik2015}.
        An early study of this problem was by Stanis\l{}aw Ulam \cite{beckenbach2013modern}
        as a toy example to illustrate the Monte Carlo method in a textbook,
        which lead to its byname \emph{Ulam's problem}. Though, it should be noted
        that in the same year Ref.~\cite{schensted1961} also discusses the
        connection of LIS to \emph{Young tableaux}.
        Ulam's study found that LIS of random permutations have a mean length
        $l$ which grows with the size of the sequence $n$ as
        $\avg{l} = c\sqrt{n}$. The Monte Carlo simulations estimated $c \approx 1.7$
        and in the years since then $c=2$ was proven rigorously \cite{aldous1999longest}.

        But the length of LIS of permutations attracted much more interest.
        In mathematics the whole distribution $p(l)$ was analyzed.
        First upper and lower tails were determined rigorously
        \cite{Seppalainen1998,logan1977variational,deuschel1999increasing},
        and later it was proven that the central part is
        a Tracy-Widom distribution \cite{baik1999distribution}. At the time this result was an unexpected
        connection between LIS and random matrix theory, where this Tracy-Widom
        distribution describes the fluctuations of the largest eigenvalues of
        the Gaussian unitary ensemble, i.e., an ensemble of Hermitian random
        matrices. In the following years it turned out that the LIS was an
        extremely simple model at the center of a growing class of seemingly unrelated
        problems. Beginning with a mapping of a 1+1-dimensional polynuclear
        growth model of the Kardar-Parisi-Zhang type onto LIS \cite{Prahofer2000universal}
        a plethora of models were shown to exhibit the properties of LIS of
        random permutations, namely that their fluctuations are distributed
        according to one of the Tracy-Widom distributions. Examples range from
        other surface growth processes, like a direct mapping of a ballistic
        deposition model on LIS \cite{majumdar2004anisotropic} or experimental
        observations of a Tracy-Widom distribution in the fluctuations of
        real surface growth \cite{takeuchi2010universal}, to the totally
        asymmetric exclusion process \cite{Johansson2000} and directed polymers \cite{Baik2000}.
        For an overview and insight into the connections between these models,
        there are some review articles \cite{Kriecherbauer2010,bouchaud2011complex,corvin2012}.

        Recently, different ensembles of sequences than the random permutation
        were studied like random walks with different distributions of their
        jump lengths \cite{angel2017increasing,mendoncca2017empirical,borjes2019large,mendonca2019asymptotic}.

        Besides its role in mathematics and physics, the LIS found applications
        in computer science, where it is suggested as a measure of
        \emph{sortedness} of large amounts of data \cite{gopalan2007} or to
        find structures in time series while preserving privacy of the data,
        which is useful in the context of, e.g., fraud detection using financial
        data streams \cite{bonomi2016differentially}.
        Also in bioinformatics the LIS found applications in the context of
        sequence alignment, e.g., for DNA or protein sequences \cite{Zhang2003}.

        Here, we are interested in another property of this famous problem.
        First, note that the LIS is not necessarily unique for any given sequence.
        For example, consider the sequence $\vec\sigma = (7, 9, 4, 1, 0, 6, 3, 8, 5, 2)$. While the
        length $l=3$ of the longest increasing subsequence is uniquely defined, this
        sequence has $M=7$ distinct LIS: $(4, 6, 8)$, $(1, 6, 8)$, $(0, 6, 8)$, $(1, 3, 8)$, $(0, 3, 8)$, $(1, 3, 5)$, $(0, 3, 5)$.
        As mentioned above, the length $l$ is thoroughly studied, but about
        the number of distinct LIS $M$ of a given sequence, only very little is
        known. Nevertheless, for example for the above mentioned application like
        determining sortedness
        or fraud detection, the actual number of distinct LIS will allow to estimate the
        reliability of decisions based on the LIS calculation much better.
        For this reason and to gain fundamental insight into the solution space structure,
        we study by computer simulations \cite{practical_guide2015} here this
        quantity,
        namely its logarithm, i.e., the entropy $S= \ln{M}$.

        One of the few results is that the number of increasing subsequences of
        a fixed length grows exponentially in $n$ \cite{hammersley1972,romik2015}.
        Although, this suggests that it is infeasible to count the LIS by
        enumeration, we will introduce in Sec.~\ref{sec:counting} an algorithm
        to count the LIS efficiently without the need to enumerate them. Due to
        the exponential growth, it indeed makes sense to finally measure the entropy $S $.
        Since we want to explore the whole distribution of the entropy, including
        the tails of extremely rare events with probabilities of, say, $10^{-100}$,
        we have to apply a sophisticated Markov chain sampling scheme, which
        will be explained in Sec.~\ref{sec:mcmc}.
        Finally, Sec.~\ref{sec:results} shows the results of our study before
        Sec.~\ref{sec:conclusions} concludes this study.
        But first, we introduce the two ensembles we studied in Sec.~\ref{sec:k}.

    \section{Models and Methods}
        For completeness we define the LIS in a more formal way than in the
        introduction. Let $\vec\sigma = (\sigma_1, \sigma_2, \ldots, \sigma_n)$ be
        a sequence of numbers.
        A LIS is a longest sequence $\lambda_{\vec\sigma} = (\sigma_{i_1}, \sigma_{i_2}, \ldots, \sigma_{i_l})$
        with $\sigma_{i_1} \le \sigma_{i_2} \le \ldots \le \sigma_{i_l}$
        such that $i_1 < i_2 < \ldots < i_l \le n$. We denote by $M$
        the number of distinct sequences
        $\lambda_{\vec\sigma}$ fulfilling this property and $S = \ln{M}$ is the entropy.

        \subsection{Ensembles of random sequences}
            \label{sec:k}

\begin{figure}[htb]
                \centering

                \subfigure[\label{fig:lis:perm}]{
                    \includegraphics[scale=1]{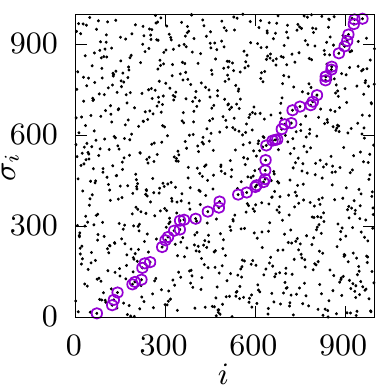}
                }
                \subfigure[\label{fig:lis:rep}]{
                    \includegraphics[scale=1]{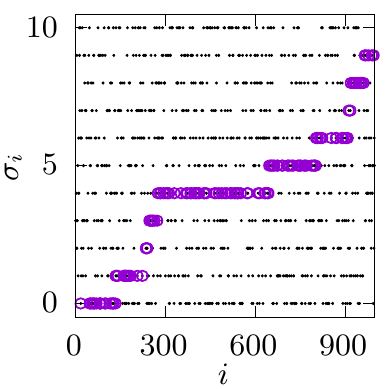}
                }

                \caption{\label{fig:lis}
                    Visualization of two sequences $\vec\sigma$. The horizontal axis
                    shows the index $i$ of the value $\sigma_i$.
                    The elements belonging to one LIS are marked by circles.
                    \subref{fig:lis:perm} Random permutation.
                    \subref{fig:lis:rep} Random sequence with $11$ distinct
                    elements.
                }
            \end{figure}

            In this study, we scrutinize two ensembles of random sequences:
            First and more in depth, random permutations, for which an example
            with one LIS marked is visualized in Fig.~\ref{fig:lis}\subref{fig:lis:perm}.

            Second, we study a parameterized random sequence consisting of
            at most $K+1$ distinct ordered elements. We call this ``$K$ ensemble''.
            An example for $K=10$ is
            shown in Fig.~\ref{fig:lis}\subref{fig:lis:rep}. In the limit $K=0$, it consists
            only of identical elements and has therefore a unique LIS with a length
            of $l=n$. The other limit $K \to \infty$ consists of sequences
            with unique elements, which can be mapped to a permutation by replacing
            each element by its rank, which in turn will not change the LIS. Thus, we
            can interpolate with $K$ between a non-degenerate LIS to the well
            known case of random permutations. Indeed, the length of the LIS
            of this ensemble was studied in Ref.~\cite{houdre2009}.

            As a technical remark, note that the algorithms explained in the
            following chapter find the strictly monotonic increasing subsequence,
            but for the $K$ ensemble, we want to find the weakly increasing
            subsequence. With a simple mapping of the sequence
            (with elements from $\mathbb{N}_0)$ to a new sequence
            $\alpha_i = \sigma_i + \frac{i}{n}$ of rational numbers
            we can apply algorithms for strict LIS on $\vec\alpha$ to find all weak LIS
            of $\vec\sigma$.

    \subsection{Counting the number of distinct LIS}
        \label{sec:counting}
        Algorithms to find the length of the LIS are rather simple and there
        exists some variety. A popular choice is patience sort \cite{Mallows1963problem}, which
        originally is a sorting algorithm especially suited for partially
        sorted data \cite{Chandramouli2014patience}, but can be simplified to an efficient algorithm to
        find the length of the LIS of a given sequence in time $\mathcal{O}(n \ln n)$ \cite{aldous1999longest}.
        But there are more alternatives, e.g., a fast algorithm in
        $\mathcal{O}(n \ln\ln n)$ \cite{Bespamyatnikh2007,Crochemore2010}, approximate algorithms
        for sequences whose members can not be saved \cite{Arlotto2015} or algorithms which are
        exact within a sliding window \cite{Albert2004}.
        Even for the enumeration of LIS, there is literature introducing algorithms \cite{Bespamyatnikh2007},
        which are able to, e.g., generate LIS with special properties \cite{Youhuan2016}.

        Here, we introduce a method to count (and enumerate) distinct LIS of
        any sequence efficiently. Note that we do not claim to be the first to
        introduce an algorithm to count the number of LIS. Some of the
        existing enumeration algorithms could be extended with the same
        principle we use to allow for efficient counting. Also there
        is at least one algorithm description for counting the LIS in a well
        known programmer forum \cite{stackoverflow}. However, we could not find
        any reference to published literature.
        Therefore we show our approach, which is an extension of patience
        sort.

        Like patience sort, this method takes elements sequentially from the
        front of the sequence and places them on top of a selected stack
        from a set $(s_1,\ldots,s_k)$ of stacks, such that
        each stack $s_i$ is sorted in decreasing order, i.e., the smallest element is on top
        of the stack, and the number of stacks is minimal. Thus, in the beginning there
        is just one stack containing the first element of the given sequence. This placement can be
        achieved by always placing the current element taken
        from the sequence on the leftmost stack,
        whose top element is larger than the current element. If this is not possible,
        i.e., if all top elements are smaller than
        the current element, one opens a new
        stack $s_{k+1}$ right of the currently rightmost stack.
        Note that therefore the top elements of the stacks
        are ascendingly sorted and the correct stack of each element can be found
        via binary search. The final number of stacks is equal to the length of
        the LIS $l$ \cite{aldous1999longest}.

\begin{figure}[htb]
            \centering

            \subfigure[]{\label{fig:dag:snapshot}
                \includegraphics[scale=1]{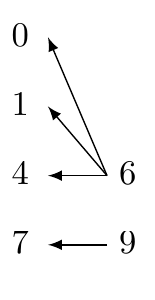}
            }\quad
            \subfigure[]{\label{fig:dag:snapshot2}
                \includegraphics[scale=1]{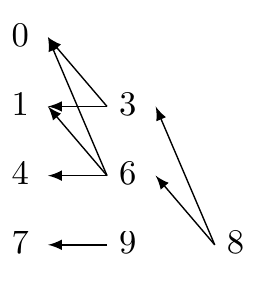}
            }\quad
            \subfigure[]{\label{fig:dag:all}
                \includegraphics[scale=1]{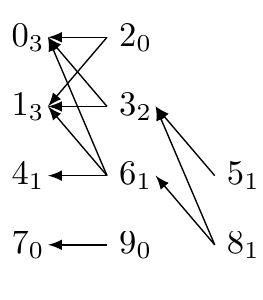}
            }

            \caption{\label{fig:dag}
                Construction of the DAG for the sequence $(7, 9, 4, 1, 0, 6, 3, 8, 5, 2)$.
                Stacks grow upwards.
                \subref{fig:dag:snapshot} Partial DAG for $(7, 9, 4, 1, 0, 6)$.
                \subref{fig:dag:snapshot2} Partial DAG for $(7, 9, 4, 1, 0, 6, 3, 8)$.
                \subref{fig:dag:all} Complete DAG with annotations (subscript)
                labeling the number of paths
                to reach the corresponding element from the rightmost stack.
                Summing the subscripts of the leftmost stack yields the total
                number of paths originating from the rightmost stack, i.e., the
                number of all LIS, here $M=7$.
            }
        \end{figure}

        For counting the LIS, we need to extend this algorithm by
        introducing pointers. The basic idea is that for any LIS, exactly one number will be
        taken from each stack \footnote{Other numbers is the same stack appear earlier in the same
        sequence but are larger, or they appear later but are smaller.}. These pointers will
        take care of the order constraints in the following way:
        Each time an element is placed on a stack, pointers
        are added to some elements of the previous stack. This idea is already described
        in Ref.~\cite{aldous1999longest}, but additionally to the pointers mentioned
        there, which will point to the currently topmost element of the previous
        stack, we also add pointers to all elements of the previous stack
        which are smaller than the current element. The meaning of such a pointer
        from any element $\sigma_{j_1}$ to $\sigma_{j_2}$ ($j_1>j_2$) will be that in a LIS
        $\sigma_{j_2}$ can appear before $\sigma_{j_1}$.
        An example structure is
        shown in Fig.~\ref{fig:dag}.

        The set of all pointers, i.e., edges, forms a directed acyclic graph (DAG).
        The DAG can be used to enumerate all LIS by following
        all paths originating from any element of the rightmost stack. This  will
        yield all LIS in reversed order. For our purpose, we just have to count
        all paths originating from the rightmost
        stack. Therefore we propagate the information by how many paths an
        element can be reached through the DAG. All elements of the rightmost
        stack are initialized with $1$. The elements of the stack left are
        assigned the sum of all incoming edges. This is repeated until the
        leftmost stack is reached. The sum of all paths ending in elements of
        the leftmost stack is the total number of LIS.

        To estimate the run time, note that we have to iterate over all incoming
        edges, of which there are at most $\mathcal{O}(n^2)$ in a DAG with $n$
        nodes. Also the construction takes the maximum of the number of edges
        for constructing the pointers and $\mathcal{O}(n\ln n)$ for constructing
        the stacks, such that the runtime of this algorithm is
        $\mathcal{O}(n^2)$ in the worst case. Note however that typical DAGs
        generated here have far fewer edges. We observed that typically the length of a LIS
        and therefore the number of stacks is $\mathcal{O}(\sqrt{n})$. Each
        stack can only be connected to the previous stack. Assuming that
        stacks are typically of size $\mathcal{O}(\sqrt{n})$, this leads to at most
        $\mathcal{O}(n)$ edges between each pair of neighboring stacks and therefore
        $\mathcal{O}(n \sqrt{n})$ total edges.

    \subsection{Sampling rare events}
        \label{sec:mcmc}

        Using the algorithm above, we can determine the number of LIS $M$ for
        arbitrary sequences. Therefore, generating random sequences allows us
        to sample $S = \ln M$, build histograms from the samples and estimate
        the distribution $p(S)$ from them. But to observe any event which occurs
        with a probability of $r$, we would have to generate $\mathcal{O}(1/r)$
        samples and $\mathcal{O}(1/r^2)$ to reduce the statistical error enough
        to determine the probability with reasonable accuracy. Since we would
        like to know the probability distribution also in the extremely rare
        event tails, we have to use a more sophisticated method than this
        proposed \emph{simple sampling}.

        Our approach is to bias the ensemble in a controlled way towards
        extremely improbable configurations, gather enough samples there
        and correct the bias afterwards. This will lead to small statistical
        errors across large parts of the support. This method \cite{Hartmann2002Sampling}
        was successfully
        applied in a wide range of problems from graph theory \cite{Hartmann2011,schawe2019large},
        over stochastic geometry \cite{schawe2018avoiding}, nonequilibrium work distributions
        \cite{Hartmann2014high},
        the Kardar-Parisi-Zhang equation \cite{kpz2018} to the exploration of the tails of
        the distribution of the LIS's length for random permutations and
        random walks \cite{borjes2019large}.

        The exact method is inspired by equilibrium thermodynamics, where the
        Metropolis algorithm \cite{metropolis1953equation} is used to generate samples of systems in the
        canonical ensemble at some \emph{temperature} $T$, which governs the
        typical values of the \emph{energies} observed in this system. Here,
        we identify the energy with our observable of interest $S$. This allows
        us to use the ``temperature'' parameter to bias the generated states
        towards improbable values of $S$.

        \begin{figure*}[htb]
            \centering

            \includegraphics[width=\linewidth]{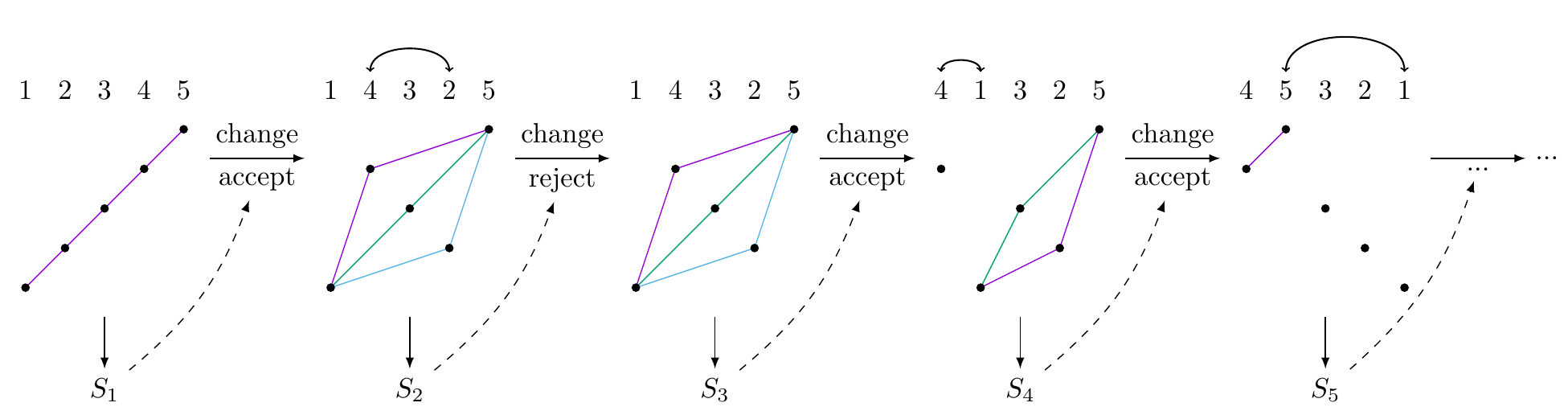}

            \caption{\label{fig:mc}
                Sketch of a Markov chain of sequence realizations generated by swaps
                of two random elements of the permutation. All distinct LIS are
                marked by lines of distinct color.
                The acceptance of a sequence as the next sequence of the Markov
                chain is dependent on the number of LIS in the realization $M$,
                since the energy is identified with $S = \ln M$.
            }
        \end{figure*}

        This method builds a Markov chain consisting of sequences $\vec\sigma^{(i)}$,
        where $i$ is the step counter of the chain. For each step in the chain,
        from the present sequence $\vec\sigma^{(i)}$ a trial sequence
        $\vec\sigma'$ is constructed by performing some changes to $\vec\sigma^{(i)}$.
        For the standard ensemble of random permutations, where we have performed
        large-deviation simulations, we used as change to swap two elements.
        The trial sequence  is accepted, i.e., $\vec\sigma^{(i+1)}=\vec\sigma'$
        with the Metropolis acceptance
        probability $P_\mathrm{acc} = \min\left(1, e^{-\Delta S/T}\right)$
        depending on the temperature $T$ and the change $\Delta S$
        between $\vec\sigma'$ and $\vec\sigma^{(i)}$.
        Otherwise the previous sequence is repeated in the chain, i.e.,
        $\vec\sigma^{(i+1)}=\vec\sigma^{(i)}$
        This procedure
        is sketched in Fig.~\ref{fig:mc} and will eventually result in
        sequences $\vec\sigma$ occurring in the chain which are distributed
        according to
        \begin{align}
            Q_T(\vec\sigma) = \frac{1}{Z_T} e^{-S(\vec\sigma)/T} Q(\vec\sigma)
        \end{align}
        where $Q(\vec\sigma)$ is the natural distribution of sequences, which we
        would obtain from simple sampling and $Z_T$ is the partition function
        of our artificial temperature ensemble. From here it is just a question
        of elemental algebra to connect our estimates of the probability density
        function in the artificial temperature ensemble $p_T(S)$ to the
        distribution of the unbiased ensemble we want to study $p(S)$:
        \begin{align}
            p_T(S) &= \sum_{\{\vec\sigma | S(\vec\sigma) = S \}} Q_T(\vec\sigma)\\
                   &= \sum_{\{\vec\sigma | S(\vec\sigma) = S \}} \frac{1}{Z_T} e^{-S(\vec\sigma)/T} Q(\vec\sigma)\\
                   &= \frac{1}{Z_T} e^{-S(\vec\sigma)/T} p(S).
        \end{align}
        Depending on the value of $T$, a simulation will generate data for $S$ in a specific
        intervall. Thus, to obtain the distribution $p(S)$ over a large range of the support,
        we performed simulations for many values of $T$.
        This requires finely tuned values of the temperatures.
        The ratio of all constants $Z_T$ can be obtained from overlaps of
        $p_{T_i}$ and $p_{T_j}$, since the actual distribution needs to be unique, i.e.,
        \begin{align}
            p_{T_j}(S) e^{S/T_j} Z_{T_j} = p_{T_i}(S) e^{S/T_i} Z_{T_i}.
        \end{align}
        We used in the
        order of 30 temperatures per size $n$, where larger sizes typically
        required more temperatures. Also, like for all Markov chain Monte Carlo
        one has to carefully ensure the equilibration of the process and
        discard sequences of the chain which are still correlated too much with previous
        sequences. Note that equilibration can be ensured rather conveniently
        \cite{Hartmann2002Sampling}, by performing two sets of simulations starting
        with very different initial sequences, with low and with high values of $S$,
        respectively. The Markov chains can be considered to be equilibrated, when
        the values of $S$ agree within fluctuations between the two sets.

    \section{Results}
        \label{sec:results}

        First, we study the behavior of typical sequences for the two ensembles.
        In the second part, we will investigate the large-deviation behavior
        of the standard permutation ensemble.

        \subsection{Typical behavior}
\begin{figure}[htb]
            \centering
            \subfigure[]{\label{fig:rp_histo}
                \includegraphics[scale=1]{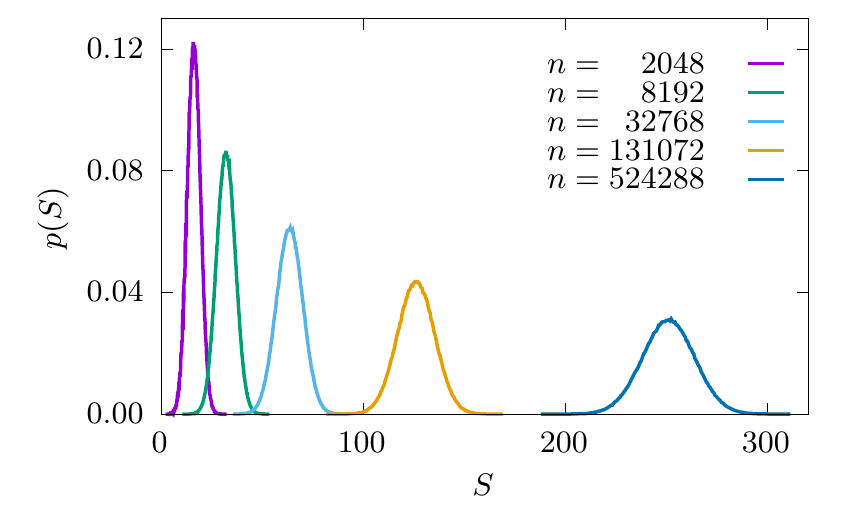}
            }
            \subfigure[]{\label{fig:rp_histo_collapse}
                \includegraphics[scale=1]{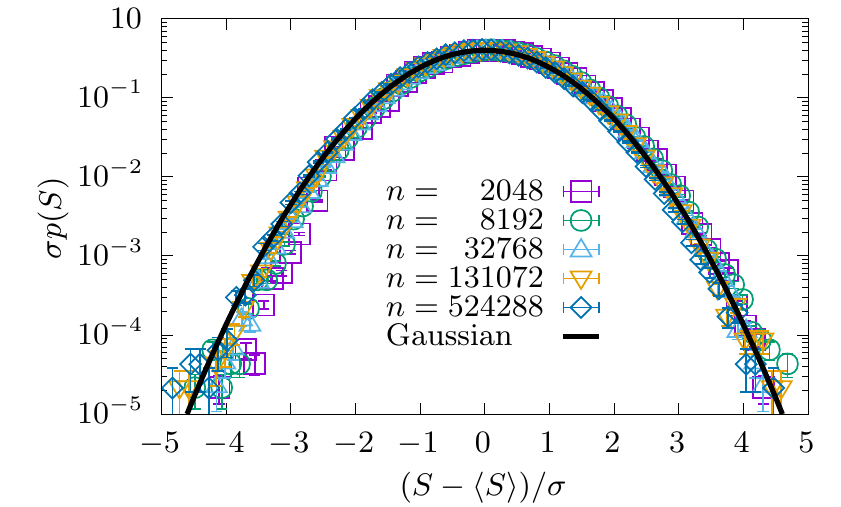}
            }
            \caption{\label{fig:rp_histos}
                \subref{fig:rp_histo}
                Probability density $p(S)$ of the entropy $S$ for
                random permutations of different length $n$, obtained with simple sampling.
                \subref{fig:rp_histo_collapse} Probability densities $p(S)$
                collapse on an approximate standard Gaussian shape for multiple system sizes
                if shifted by their mean $\avg{S} \approx 0.347 \sqrt{n}$ and scaled
                with their width $\sigma \approx 0.49 \sqrt[4]{n}$.
                Note the logarithmic vertical axis.
            }
        \end{figure}

        To investigate the typical behavior of the permutation ensemble, we consider
        different system sizes up to large sequences of $n=524288=2^{19}$ elements.
        The estimated probability density functions $p(S)$ of the LIS-entropy
        of permutations is shown in the range of typical probabilities
        in Fig.~\ref{fig:rp_histos}\subref{fig:rp_histo}.
        These data are collected over $10^6$ samples for each system size.
        Clearly, the mean value and width of the distribution increase with
        $n$.

        Indeed, we observe for the mean a growth of the form
        \begin{align}
            \label{eq:mean}
            \avg{S} = c \sqrt{n}.
        \end{align}
        (Also see below in Fig.~\ref{fig:k_mean}\subref{fig:k_p_mean}.)
        Note that the fits resulted in rather
        large reduced $\chi^2$ goodness of fit values (caused by the very high
        precision of the measured means) suggesting that there are
        corrections to this form for finite sizes. Our best estimate for the
        prefactor under the assumption that the above relation is correct,
        is $c \approx 0.347$. Also, for the standard deviation we observe
        a similar simple relation of $\sigma_S = b \sqrt[4]{n}$ with $b\approx 0.49$.

        We notice that the growth of the mean entropy $\avg{S} \approx c \sqrt{n}$
        and of the mean count $\avg{M} \approx e^{c_2 \sqrt{n}}$ with $c_2 \approx 0.44$ (not shown)
        estimated from our data, follow the same behavior as the mean number of
        increasing subsequences (IS) of length $2\sqrt{n} = \avg{l}$ \cite[Eq.~(11.5)]{hammersley1972},
        $\avg{m} \approx \left(\frac{e}{2}\right)^{4\sqrt{n}} = e^{4 \ln(e/2) \sqrt{n}}$.
        Thus, since $c_2 < 4 \ln(e/2) \approx 1.2$, our numerical results suggest
        the actual mean count of LIS to be much
        lower (and $e^{\avg{S}}$ even lower in accordance to Jensen's inequality).
        In other words, the number of \emph{longest} increasing subsequences
        is exponential lower than the number of increasing subsequences
        of the same length. This can be understood in the following way:
        We consider IS of given length $\overline{l}$, which is the average LIS length for
        this value of $n$. Now, looking at the ensemble of sequences,
        some will have a LIS length $l<\overline{l}$. For them, there are no IS of
        length $\overline{l}$, so they will not contribute any IS to the average.
        This will be a fraction of sequences. Some sequences will have LIS length
        of $l=\overline{l}$, they
        will contribute all their LISs to the average. Finally, some fraction
        of sequences will have a LIS length $l>\overline{l}$. Here, all
        subsequences of length $\overline{l}$ of all LIS will be IS contributing
        to the average. Since there are exponentially many subsequences
        (and maybe even more IS which are not subsequences of a LIS),
        they will dominate the average number of IS, thus leading to a stronger
        exponential growth as compared to the average number of LIS.

        We use our estimates for the mean and standard deviation to rescale
        the distributions for different system sizes in
        Fig.~\ref{fig:rp_histos}\subref{fig:rp_histo_collapse} and observe
        a collapse on a shape which can be approximated well by a standard Gaussian.
        Especially, the strongest deviations from this scaling form occur
        for small sizes, while the larger sizes seem to converge to the limiting
        shape. This is expected since the corrections to the scaling we used,
        which are mentioned above, should be stronger for smaller values of $n$.
        We backed this observation by classical normality
        tests \cite{Agostino1973,press2007numerical,scipy},
        which are able to distinguish this distribution from a normal distribution
        with very high confidence at small system sizes, but become less confident
        for the largest system sizes (details not shown here).
        Especially the weak Kolmogorov-Smirnov test
        is not able to distinguish the distributions from a normal distribution
        with a significance level below $10\%$ for all sizes $n\ge 65536$ of for
        our sample of $10^6$ realizations.

        Due to our limited sample size, the tails of the measured distribution
        are subject to statistical errors. In Sec.~\ref{sec:fartails} we will
        present higher quality data for the far tails to show that the
        approximation
        by a Gaussian shape is valid deep into the tails. Even for extreme
        rare events we can not exclude the possibility that the distribution
        converges to a Gaussian in the large-$n$ limit.

        Next, we look at the ensemble of random sequences with a limited number
        of $K+1$ distinct elements. For constant values of $K$ Fig.~\ref{fig:k_mean}\subref{fig:k_c_mean}
        shows the average entropy. The trivial case of $K=0$, which allows only
        one LIS of length $l=n$, corresponds to an entropy of $S = 0$ and is not
        visualized. The case $K=1$, which consists of sequences containing two
        distinct elements, has a low and interestingly almost $n$-independent
        entropy. There are typically only one or two distinct LIS in such sequences
        independent of the length of the sequence. Our data for larger values of $K$
        show two phenomena. First, larger values of $K$ lead to larger entropies
        and second, the dependency of the entropy on the length of the sequence
        diminishes for the limit of large $n$, i.e., for each fixed $K$ there
        should be a limit entropy approached for $n\to\infty$. Indeed, fits
        of a function $f(n) = b+\frac{a}{\ln(n-c)}$
        with the limiting value $\lim_{n\to\infty} f(n) = b$ to our data confirm this guess
        for all values of $K$ we considered. Note that the shape of the fitting form is
        purely heuristical. We first tried standard shapes like approaching a constant
        with a power law or an exponential, but they did not work out well.

        Since $b$ seems to grow roughly linear with $K$ (not shown),
        this leads to the conjecture that for $K \propto n$ the saturation
        of the entropy should not occur and instead grow with the size $n$.
        Especially, we observed this behavior for the permutation case, which
        is identical to the $K \to \infty$ limit, as explained in Sec.~\ref{sec:k}.
        Indeed, in Fig.~\ref{fig:k_mean}\subref{fig:k_p_mean} we can observe a quick convergence
        with increasing $n$ to the behavior of the random permutation. Our conjecture
        Eq.~\eqref{eq:mean} for its growth is visualized as a line.

        \begin{figure}[htb]
            \centering
            \subfigure[]{\label{fig:k_c_mean}
                \includegraphics[scale=1]{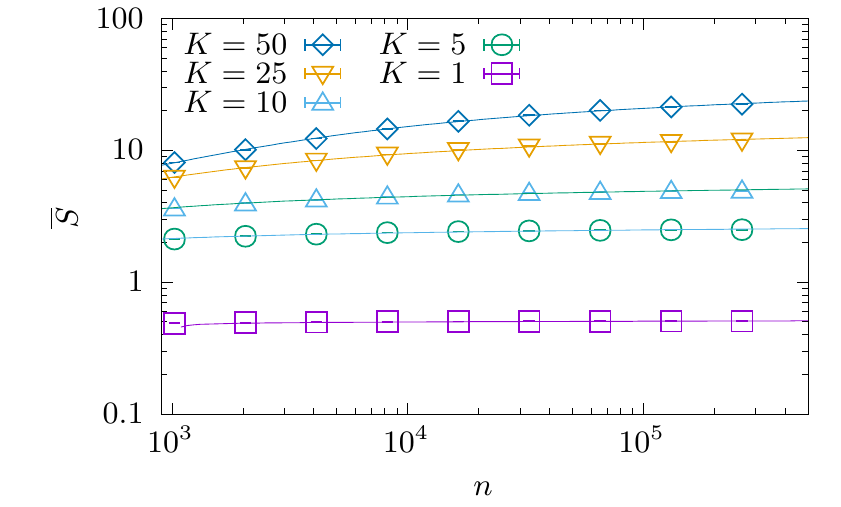}
            }
            \subfigure[]{\label{fig:k_p_mean}
                \includegraphics[scale=1]{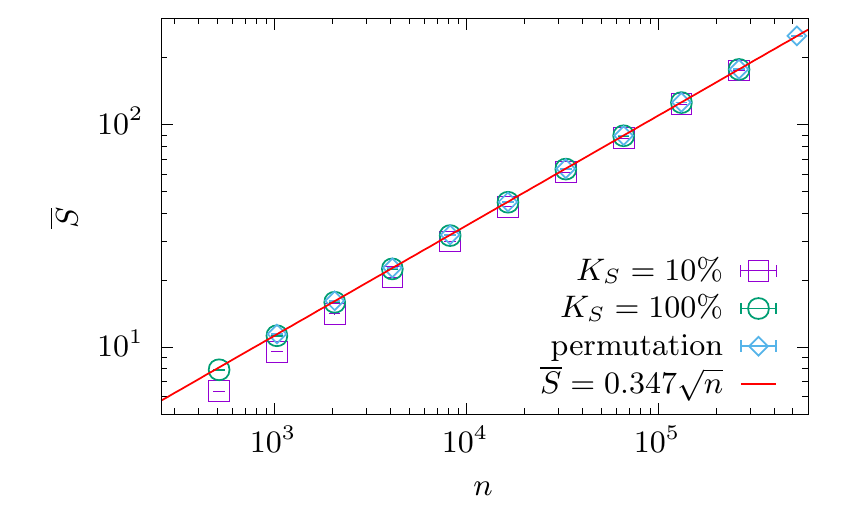}
            }
            \caption{\label{fig:k_mean}
                Average entropy $\overline{S}$ as a function of the sequence
                length $n$, for the permutation ensemble and for
                the $K$ ensemble with different values of $K$.
                Error bars are usually smaller than the width of the lines.
                \subref{fig:k_c_mean} $K$ ensemble with constant $K$.
                Lines are fits of the form $f(n) = b+\frac{a}{\ln(n-c)}$
                \subref{fig:k_p_mean} Permutation ensemble
                and  with $K \propto n$.
                The line is the growth observed for random permutations $\avg{S} \approx 0.347 \sqrt{n}$.
            }
        \end{figure}

    \subsection{The Far Tails}
    \label{sec:fartails}

        In this section we study the distribution of the entropy of the LIS
        for the random permutation ensemble with a focus on the far tails.
        Due to the much larger numerical effort, we are able to show results up
        to sequence lengths of $n=8192$.

\begin{figure}[htb]
            \centering
            \subfigure[]{\label{fig:rp_ld_histo}
                \includegraphics{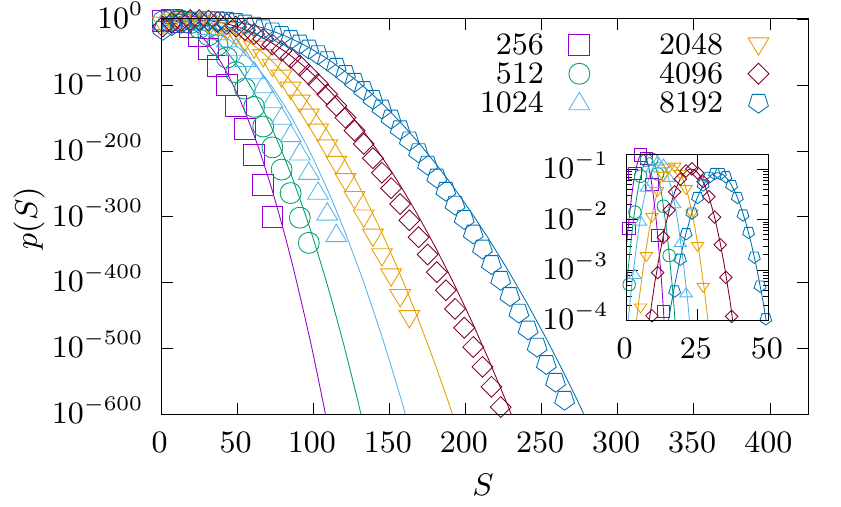}
            }
            \subfigure[]{\label{fig:rp_ld_histo_collapse}
                \includegraphics[scale=1]{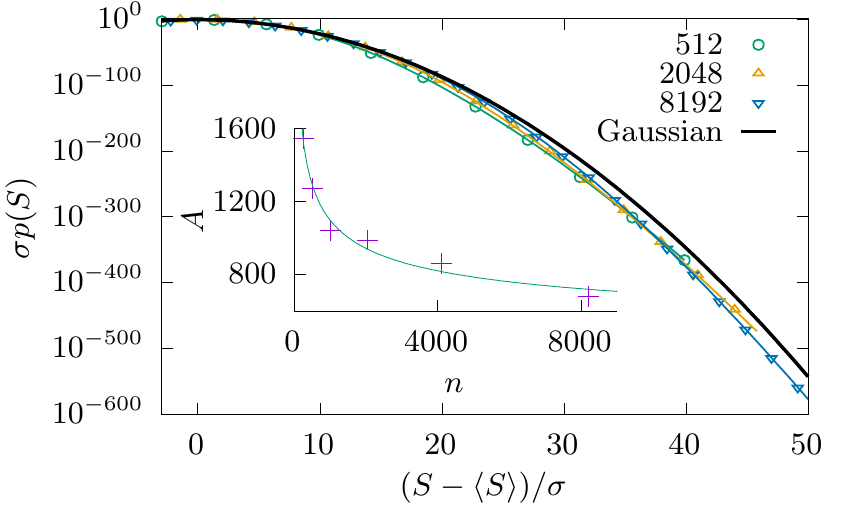}
            }

            \caption{\label{fig:rp_ld_histos}
                \subref{fig:rp_ld_histo} Probability density $p(S)$ for multiple system sizes with extremely
                high precision data for the far tails. The inset shows a
                zoom on the high probability region. The lines are fits to Gaussian
                distributions, which fit very well in the high-probability region,
                but do not describe the whole tails of the distribution.
                \subref{fig:rp_ld_histo_collapse} Same rescaling of the axes as
                Fig.~\ref{fig:rp_histo_collapse}. This shows that the different
                system sizes move towards the Gaussian for larger sizes.
                The lines are linear interpolations of all available data points,
                not all of them are shown as symbols for clarity.
                The inset shows the area between the logarithm of the rescaled
                distribution and the logarithm of a standard normal distribution
                with a fit used for extrapolation.
            }
        \end{figure}

        The data presented in the previous section, which was
        obtained via simple sampling, resulted in a distribution which
        appeared to be very well approximated by a Gaussian. We want to investigate
        whether this is still true when including our
        high-precision estimates of the far tails. Here, we can observe a slightly
        faster than Gaussian decay, see
        Fig.~\ref{fig:rp_ld_histos}\subref{fig:rp_ld_histo}. There the
        distribution for a selection of sequence lengths $n$ is shown including
        the far tail and fitted with (normalized) Gaussians. While they describe
        the high-probability region of the distribution (shown in the inset)
        very well, the deviation becomes stronger for increasingly rare events.

        To test whether this deviation remains in the $n\to\infty$ limit, we
        rescale the distributions, like in Fig.~\ref{fig:rp_histos}\subref{fig:rp_histo_collapse}, to
        be independent of system size, see Fig.~\ref{fig:rp_ld_histos}\subref{fig:rp_ld_histo_collapse}.
        First, we see that this collapse does not
        work as well in the far tails, as it does in the high-probability region.
        Interestingly, there is a crossing for different system sizes. Left of
        the crossing larger sizes tend towards the Gaussian, which hints that
        for larger sizes the Gaussian approximation becomes better even in the
        intermediate tails. Careful examination of the crossing shows that its
        position is dependent on the system size and larger systems cross farther
        on the right than smaller systems. This is again a hint that the Gaussian
        approximation becomes valid over larger ranges of the distribution for
        larger system sizes.

        To quantify this observation, we can not use classical statistical tests
        like we could do it for the data we obtained via simple sampling.
        Instead we will
        use a crude estimate of similarity, similar to one already used in \cite{schawe2020large}.
        We compare the area $A$ between the logarithm of the scaled empirical
        distributions $p_s(x) = \sigma p(S)$, where $x = (S-\avg{S})/\sigma$ (cf.~Fig.~\ref{fig:rp_ld_histos}\subref{fig:rp_ld_histo_collapse}),
        and the logarithm of a standard normal probability density
        function $p_\mathrm{G}$
        to estimate whether they become more similar for larger sizes.
        To be comparable across all system sizes, we limit this difference to
        the largest range of the horizontal axis, for which we have data for
        all sizes, i.e.,
        \begin{align}
            A = \int_0^{40} \abs{ \ln p_s(x) - \ln p_\mathrm{G}(x) }\, \mathrm{d}x.
        \end{align}

        Using this method we observe a strictly decreasing area, as shown in the
        inset of Fig.~\ref{fig:rp_ld_histos}\subref{fig:rp_ld_histo_collapse}. If we
        extrapolate it using a power law with offset $A(n) = c + an^{-b}$ we
        obtain a result for the offset $c = 344 \pm 360$, which is within
        errorbars consistent with an offset of $0$, i.e., consistent with a
        convergence to a Gaussian. However, since
        the constants of our scaling assumption are tainted
        with hard to quantify errors, this should be interpreted as a trend and
        not as a rigorous result.
        Nevertheless this result means that we can not exclude the possibility
        that the distribution will be Gaussian in the
        right tail in the limit of $n \to \infty$.

        Next, we can use our empirical data of the distribution to test whether
        a \emph{large deviation principle} holds, that is whether the behavior
        of the distribution can be expressed by a \emph{rate function} $\Phi$ in the
        $n\to\infty$ limit, defined by $\Phi(s) = -\lim_{n\to\infty} \frac{1}{n}
        \ln p_n (sS_{\mathrm{max},n})$ \cite{Touchette2009large}, where $S_{\mathrm{max},n}$ is the maximum
        possible value for a given value of $n$ and therefore $s\in [0,1]$.
        This rescaling with the maximum value, i.e., $s=S/S_{\mathrm{max},n}$,
        is done to describe the largest fluctuations by one size-independent function
        $\Phi(s)$.
        Since we have the distributions $p_n(S)$ for multiple finite $n$,
        we can calculate \emph{empirical rate functions} $\Phi_n(s)$ for each
        $n$ and extrapolate whether they converge to a limiting curve, which
        is a strong hint that this curve is the size-independent rate function
        $\Phi(s)$, which would establish a large deviation principle.
        If a rate function exists, it governs the fluctuations around the mean.
        For example, the existence of a rate function with mild properties implies
        the law of large numbers and the central limit theorem for the corresponding
        process.
        For brevity, we will omit $n$-subscripts for $S_\mathrm{max,n}$ and $p_n(S)$.

        To analyze our data, we have to determine $S_{\mathrm{max}}$ for the LIS.
        A maximum entropy is achieved if for many groups of elements, one can
        choose independently between different elements. Thus,
        consider a sequence, which consists of groups of
        $k$ decreasing elements, followed by $k$ decreasing elements, which are
        larger than all elements before and so on. An example for $n=9$ and $k=3$
        is $(3,2,1,6,5,4,9,8,7)$. In this case a LIS would
        have length $n/k$ and can contain for each block of $k$ an arbitrary element,
        resulting in $M = k^{n/k}$ distinct LIS. The entropy
        $S= \ln M = \frac{n}{k} \ln k$ is maximized at $k=e$, and since we are
        limited to integer $k$, at $k=3$. This results in a maximum entropy of
        $S_\mathrm{max} = n \ln 3 /3$, i.e., linear in $n$.

\begin{figure}[htb]
            \centering
            \subfigure[]{\label{fig:rp_rate:usual}
                \includegraphics{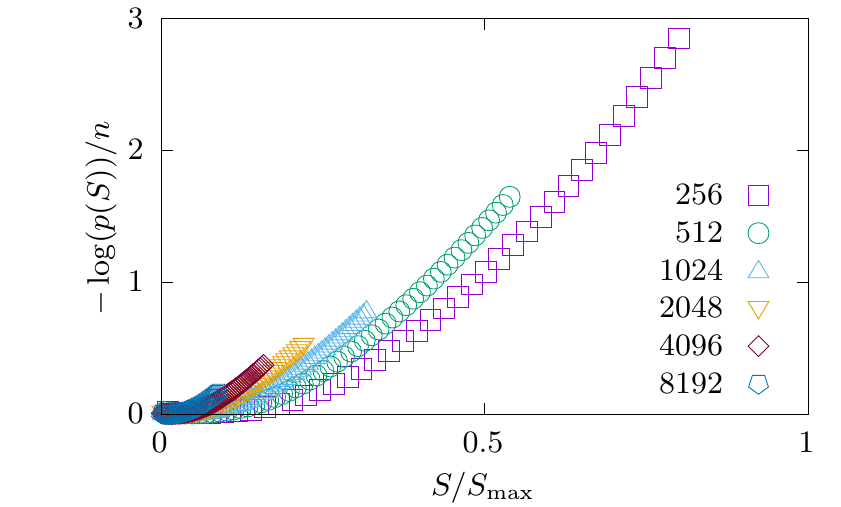}
            }
            \subfigure[]{\label{fig:rp_rate:unusual}
                \includegraphics[scale=1]{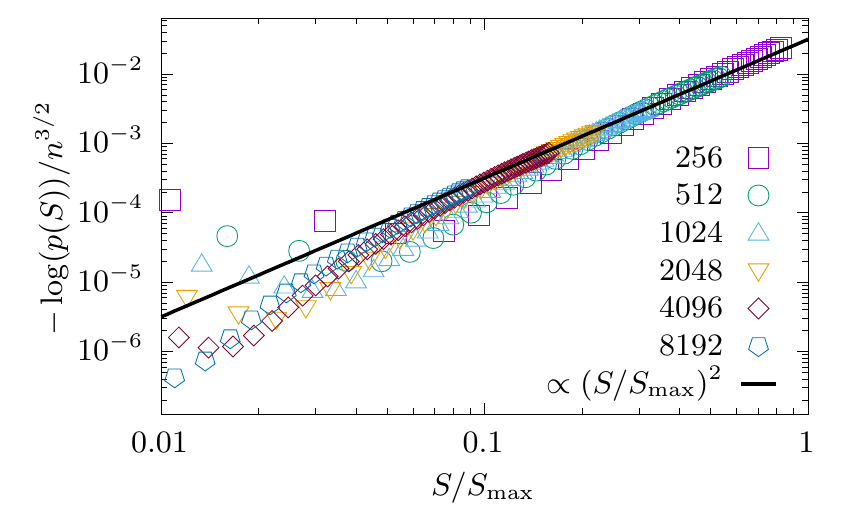}
            }
            \caption{\label{fig:rp_rate}
                \subref{fig:rp_rate:usual} Usual empirical rate function
                $\Phi(s) = -\lim_{n\to\infty} \frac{1}{n} \ln p(S)$.
                No convergence is visible, the curves shift to the left with
                increasing sequence length $n$.
                \subref{fig:rp_rate:unusual} Empirical rate function with unusual exponent
                $\Phi_u(s) = -\lim_{n\to\infty} \frac{1}{n^{3/2}} \ln p(S)$
                for the random permutation case in log-log scale to emphasize
                the convergence to a common tail with a power-law shape.
            }
        \end{figure}

        In Fig.~\ref{fig:rp_rate}\subref{fig:rp_rate:usual} the empirical rate functions are visualized,
        but no convergence to a common tail is visible. The best common tail
        we can generate, happens for a slightly modified rate function
        with an unusual exponent $\Phi_u(s) = -\lim_{n\to\infty} \frac{1}{n^{3/2}} \ln p(S)$.
        Note however, that such an exponent is not out of the question. For
        example, the rate function of the right tail of the distribution for the
        rescaled length $\tilde{l}=l/\sqrt{n}$ of LIS behaves as
        $\Phi(\tilde{l}) = -\lim_{n\to\infty} \frac{1}{n^{1/2}} \ln p(l)$ \cite{deuschel1999increasing}.
        Our result is shown in Fig.~\ref{fig:rp_rate}\subref{fig:rp_rate:unusual} in double
        logarithmic scale to emphasize the collapse in the right tail on a
        power law $\sim s^\kappa$ with a slope of $\kappa \approx 2$,
        consistent with the before observed almost Gaussian right tail.

\begin{figure}[htb]
            \centering
            \includegraphics{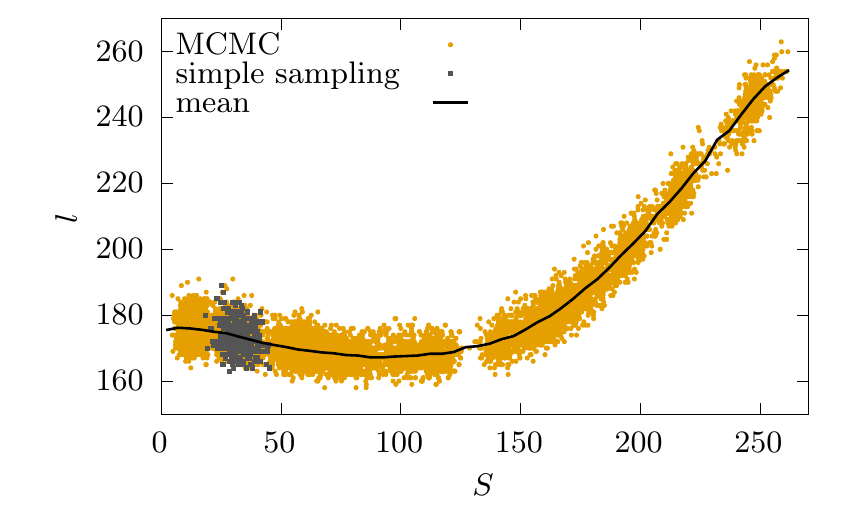}
            \caption{\label{fig:scatter8192}
                The length of the LIS $l$ as function of
                the entropy $S$ for a sequence length $n=8192$.
                The dark gray data points are gathered using
                simple sampling and represent typical sequences. The other
                data points are collected with the Markov chain Monte Carlo (MCMC)
                method described in Sec.~\ref{sec:mcmc} and represent extremely
                rare sequences with atypical entropy.
            }
        \end{figure}

        Finally, we want to understand what leads to sequences with atypically many or
        few distinct LIS. For this purpose we used the sequences generated by simple
        sampling and by the large-deviation approach
        to study their correlation with the length of the
        corresponding LIS. This might give insight into qualitative mechanisms
        governing the degeneracy of the LIS. This correlation is visualized in
        Fig.~\ref{fig:scatter8192} for random permutations of length $n=8192$.
        Typical permutations have a LIS-entropy around $\avg{S} \approx 35$
        with a typical length of $\avg{l} \approx 180$, marked by darker gray
        points. Apparently, the degeneracy of the LIS is uncorrelated with its
        length for low and intermediate values of $S$. Though, extremely
        degenerate LIS necessitate longer LIS. This is somewhat counter intuitive,
        since there are more increasing subsequences of shorter length \cite{romik2015}.
        We assume therefore that the mechanism here is that these rare sequences
        have a structure which is in some sense modular (cf.~Sec.~\ref{sec:fartails} for the
        configuration of maximum entropy) to allow for many almost
        identical LIS. These may differ independently in many places and therefore
        can combinatorically combine such small differences. Then, since a
        longer LIS has more members, the combinatorial character leads to an
        entropy advantage for long LIS. This higher
        number of combinatorial possibilities becomes necessary at some point
        to support even more degenerate LIS, thus we see a very strong
        correlation for extremely high values of $S$. However, we assume that
        for very long LIS, the entropy has to decrease again. In the extreme
        case of $l=n$ the sequence has to be sorted and can only contain a
        single LIS. Since we do not observe very long LIS in our sampled data,
        it means that they are combinatorially suppressed. In order to have access to this
        region, one would have to perform a biased sampling with respect to the length and
        measure the entropy, or, even better, measure the two-dimensional distribution
        $p(S,l)$ by a two-temperature large-deviation approach, which would be numerically
        very demanding.

    \section{Conclusions}
    \label{sec:conclusions}
        Here, we studied the entropy $S$ of longest increasing subsequences of
        random permutations by counting the number of distinct LIS. Using an
        extension of the patience-sort algorithm, this can be readily obtained
        for any given sequence. Especially
        we applied Markov chain Monte Carlo techniques to explore the far tail
        of the  probability distribution of $S$ in the regime of extremely rare events
        with probabilities less than $10^{-600}$.

        Concerning the typical behavior, we found that the average entropy
        grows as a square root in the length of the permutation, i.e., the
        number of LIS grows exponentially, as expected. The fluctuations of
        the entropy are in good approximation Gaussian, but show deviations
        from this shape in the far tails.

        Further, we use the data of the far tails to empirically scrutinize
        the rate function, the central piece of large-deviation theory. For the
        right tails we propose a rate function with an unusual exponent
        $\Phi\left(S/S_\mathrm{max}\right) = -\lim_{n\to\infty} \ln p\left(S\right) / n^{3/2} \sim \left(S/S_\mathrm{max}\right)^2$,
        towards which the right tails of all studied system sizes seem to
        converge. This means, the standard large-deviation principle, where one
        would see a convergence with
        a factor $1/n$ instead of $1/n^{3/2}$, does not hold, but still the tails
        of the distribution can be described by some rate function. Note that
        for the distribution of LIS lengths also a rate function with a factor
        different from $1/n$ was found in previous work.

        Additionally to random permutations, we studied an ensemble with a limited
        amount of distinct elements in the sequences. For fixed number of
        distinct elements, we observed that $S$ converges to a constant value which is
        independent of the sequence length for long sequences. For any number
        of distinct symbols, which is proportional to the sequence length,
        this will converge to the same LIS-entropy as random permutations for
        large system sizes.

        Also, the datastructure used to count the LIS can be used to perform
        unbiased sampling of all LIS, which is a line of research the authors
        are working on right now. Here, also other ensembles of sequences could be of
        interest, like one-dimensional random walks.
        Finally, for future research it could be
        interesting, yet numerically extremely demanding, to study the two-dimensional
        distributions like $p(S,l)$.

    \section*{Acknowledgments}
        HS would like to thank Naftali Smith for interesting discussions
        about LIS and acknowledges the OpLaDyn grant obtained in the 4th round
        of the TransAtaltic program Digging into Data Challenge (2016-147 ANR OPLADYN TAP-DD2016).
        The simulations were performed at the HPC Cluster CARL, located at
        the University of Oldenburg (Germany) and funded by the DFG through
        its Major Research Instrumentation Programme (INST 184/108-1 FUGG)
        and the Ministry of Science and Culture (MWK) of the Lower Saxony
        State.

    \bibliography{lit}

\begin{thebibliography}{47}%
\makeatletter
\providecommand \@ifxundefined [1]{%
 \@ifx{#1\undefined}
}%
\providecommand \@ifnum [1]{%
 \ifnum #1\expandafter \@firstoftwo
 \else \expandafter \@secondoftwo
 \fi
}%
\providecommand \@ifx [1]{%
 \ifx #1\expandafter \@firstoftwo
 \else \expandafter \@secondoftwo
 \fi
}%
\providecommand \natexlab [1]{#1}%
\providecommand \enquote  [1]{``#1''}%
\providecommand \bibnamefont  [1]{#1}%
\providecommand \bibfnamefont [1]{#1}%
\providecommand \citenamefont [1]{#1}%
\providecommand \href@noop [0]{\@secondoftwo}%
\providecommand \href [0]{\begingroup \@sanitize@url \@href}%
\providecommand \@href[1]{\@@startlink{#1}\@@href}%
\providecommand \@@href[1]{\endgroup#1\@@endlink}%
\providecommand \@sanitize@url [0]{\catcode `\\12\catcode `\$12\catcode
  `\&12\catcode `\#12\catcode `\^12\catcode `\_12\catcode `\%12\relax}%
\providecommand \@@startlink[1]{}%
\providecommand \@@endlink[0]{}%
\providecommand \url  [0]{\begingroup\@sanitize@url \@url }%
\providecommand \@url [1]{\endgroup\@href {#1}{\urlprefix }}%
\providecommand \urlprefix  [0]{URL }%
\providecommand \Eprint [0]{\href }%
\providecommand \doibase [0]{http://dx.doi.org/}%
\providecommand \selectlanguage [0]{\@gobble}%
\providecommand \bibinfo  [0]{\@secondoftwo}%
\providecommand \bibfield  [0]{\@secondoftwo}%
\providecommand \translation [1]{[#1]}%
\providecommand \BibitemOpen [0]{}%
\providecommand \bibitemStop [0]{}%
\providecommand \bibitemNoStop [0]{.\EOS\space}%
\providecommand \EOS [0]{\spacefactor3000\relax}%
\providecommand \BibitemShut  [1]{\csname bibitem#1\endcsname}%
\let\auto@bib@innerbib\@empty
\bibitem [{\citenamefont {Romik}(2015)}]{romik2015}%
  \BibitemOpen
  \bibfield  {author} {\bibinfo {author} {\bibfnamefont {D.}~\bibnamefont
  {Romik}},\ }\href@noop {} {\emph {\bibinfo {title} {The Surprising
  Mathematics of Longest Increasing Subsequences}}}\ (\bibinfo  {publisher}
  {Cambridge University Press},\ \bibinfo {address} {USA},\ \bibinfo {year}
  {2015})\BibitemShut {NoStop}%
\bibitem [{\citenamefont {Ulam}(2013)}]{beckenbach2013modern}%
  \BibitemOpen
  \bibfield  {author} {\bibinfo {author} {\bibfnamefont {S.~M.}\ \bibnamefont
  {Ulam}},\ }in\ \href@noop {} {\emph {\bibinfo {booktitle} {Modern Mathematics
  for the Engineer: Second Series}}},\ \bibinfo {series and number} {Dover
  Books on Engineering Series},\ \bibinfo {editor} {edited by\ \bibinfo
  {editor} {\bibfnamefont {E.}~\bibnamefont {Beckenbach}}\ and\ \bibinfo
  {editor} {\bibfnamefont {M.}~\bibnamefont {Hestenes}}}\ (\bibinfo
  {publisher} {Dover Publications, Incorporated},\ \bibinfo {year} {2013})\
  Chap.~\bibinfo {chapter} {11}, pp.\ \bibinfo {pages} {261--281}\BibitemShut
  {NoStop}%
\bibitem [{\citenamefont {Schensted}(1961)}]{schensted1961}%
  \BibitemOpen
  \bibfield  {author} {\bibinfo {author} {\bibfnamefont {C.}~\bibnamefont
  {Schensted}},\ }\href {\doibase 10.4153/CJM-1961-015-3} {\bibfield  {journal}
  {\bibinfo  {journal} {Canadian Journal of Mathematics}\ }\textbf {\bibinfo
  {volume} {13}},\ \bibinfo {pages} {179–191} (\bibinfo {year}
  {1961})}\BibitemShut {NoStop}%
\bibitem [{\citenamefont {Aldous}\ and\ \citenamefont
  {Diaconis}(1999)}]{aldous1999longest}%
  \BibitemOpen
  \bibfield  {author} {\bibinfo {author} {\bibfnamefont {D.}~\bibnamefont
  {Aldous}}\ and\ \bibinfo {author} {\bibfnamefont {P.}~\bibnamefont
  {Diaconis}},\ }\href@noop {} {\bibfield  {journal} {\bibinfo  {journal}
  {Bulletin of the American Mathematical Society}\ }\textbf {\bibinfo {volume}
  {36}},\ \bibinfo {pages} {413} (\bibinfo {year} {1999})}\BibitemShut
  {NoStop}%
\bibitem [{\citenamefont {Sepp{\"a}l{\"a}inen}(1998)}]{Seppalainen1998}%
  \BibitemOpen
  \bibfield  {author} {\bibinfo {author} {\bibfnamefont {T.}~\bibnamefont
  {Sepp{\"a}l{\"a}inen}},\ }\href {\doibase 10.1007/s004400050188} {\bibfield
  {journal} {\bibinfo  {journal} {Probability Theory and Related Fields}\
  }\textbf {\bibinfo {volume} {112}},\ \bibinfo {pages} {221} (\bibinfo {year}
  {1998})}\BibitemShut {NoStop}%
\bibitem [{\citenamefont {Logan}\ and\ \citenamefont
  {Shepp}(1977)}]{logan1977variational}%
  \BibitemOpen
  \bibfield  {author} {\bibinfo {author} {\bibfnamefont {B.~F.}\ \bibnamefont
  {Logan}}\ and\ \bibinfo {author} {\bibfnamefont {L.~A.}\ \bibnamefont
  {Shepp}},\ }\href@noop {} {\bibfield  {journal} {\bibinfo  {journal}
  {Advances in Mathmatics}\ }\textbf {\bibinfo {volume} {26}},\ \bibinfo
  {pages} {206} (\bibinfo {year} {1977})}\BibitemShut {NoStop}%
\bibitem [{\citenamefont {Deuschel}\ and\ \citenamefont
  {Zeitouni}(1999)}]{deuschel1999increasing}%
  \BibitemOpen
  \bibfield  {author} {\bibinfo {author} {\bibfnamefont {J.-D.}\ \bibnamefont
  {Deuschel}}\ and\ \bibinfo {author} {\bibfnamefont {O.}~\bibnamefont
  {Zeitouni}},\ }\href@noop {} {\bibfield  {journal} {\bibinfo  {journal}
  {Combinatorics, Probability and Computing}\ }\textbf {\bibinfo {volume}
  {8}},\ \bibinfo {pages} {247} (\bibinfo {year} {1999})}\BibitemShut {NoStop}%
\bibitem [{\citenamefont {Baik}\ \emph {et~al.}(1999)\citenamefont {Baik},
  \citenamefont {Deift},\ and\ \citenamefont
  {Johansson}}]{baik1999distribution}%
  \BibitemOpen
  \bibfield  {author} {\bibinfo {author} {\bibfnamefont {J.}~\bibnamefont
  {Baik}}, \bibinfo {author} {\bibfnamefont {P.}~\bibnamefont {Deift}}, \ and\
  \bibinfo {author} {\bibfnamefont {K.}~\bibnamefont {Johansson}},\ }\href@noop
  {} {\bibfield  {journal} {\bibinfo  {journal} {Journal of the American
  Mathematical Society}\ }\textbf {\bibinfo {volume} {12}},\ \bibinfo {pages}
  {1119} (\bibinfo {year} {1999})}\BibitemShut {NoStop}%
\bibitem [{\citenamefont {Pr\"ahofer}\ and\ \citenamefont
  {Spohn}(2000)}]{Prahofer2000universal}%
  \BibitemOpen
  \bibfield  {author} {\bibinfo {author} {\bibfnamefont {M.}~\bibnamefont
  {Pr\"ahofer}}\ and\ \bibinfo {author} {\bibfnamefont {H.}~\bibnamefont
  {Spohn}},\ }\href {\doibase 10.1103/PhysRevLett.84.4882} {\bibfield
  {journal} {\bibinfo  {journal} {Phys. Rev. Lett.}\ }\textbf {\bibinfo
  {volume} {84}},\ \bibinfo {pages} {4882} (\bibinfo {year}
  {2000})}\BibitemShut {NoStop}%
\bibitem [{\citenamefont {Majumdar}\ and\ \citenamefont
  {Nechaev}(2004)}]{majumdar2004anisotropic}%
  \BibitemOpen
  \bibfield  {author} {\bibinfo {author} {\bibfnamefont {S.~N.}\ \bibnamefont
  {Majumdar}}\ and\ \bibinfo {author} {\bibfnamefont {S.}~\bibnamefont
  {Nechaev}},\ }\href@noop {} {\bibfield  {journal} {\bibinfo  {journal}
  {Physical Review E}\ }\textbf {\bibinfo {volume} {69}},\ \bibinfo {pages}
  {011103} (\bibinfo {year} {2004})}\BibitemShut {NoStop}%
\bibitem [{\citenamefont {Takeuchi}\ and\ \citenamefont
  {Sano}(2010)}]{takeuchi2010universal}%
  \BibitemOpen
  \bibfield  {author} {\bibinfo {author} {\bibfnamefont {K.~A.}\ \bibnamefont
  {Takeuchi}}\ and\ \bibinfo {author} {\bibfnamefont {M.}~\bibnamefont
  {Sano}},\ }\href {\doibase 10.1103/PhysRevLett.104.230601} {\bibfield
  {journal} {\bibinfo  {journal} {Phys. Rev. Lett.}\ }\textbf {\bibinfo
  {volume} {104}},\ \bibinfo {pages} {230601} (\bibinfo {year}
  {2010})}\BibitemShut {NoStop}%
\bibitem [{\citenamefont {Johansson}(2000)}]{Johansson2000}%
  \BibitemOpen
  \bibfield  {author} {\bibinfo {author} {\bibfnamefont {K.}~\bibnamefont
  {Johansson}},\ }\href {\doibase 10.1007/s002200050027} {\bibfield  {journal}
  {\bibinfo  {journal} {Communications in Mathematical Physics}\ }\textbf
  {\bibinfo {volume} {209}},\ \bibinfo {pages} {437} (\bibinfo {year}
  {2000})}\BibitemShut {NoStop}%
\bibitem [{\citenamefont {Baik}\ and\ \citenamefont {Rains}(2000)}]{Baik2000}%
  \BibitemOpen
  \bibfield  {author} {\bibinfo {author} {\bibfnamefont {J.}~\bibnamefont
  {Baik}}\ and\ \bibinfo {author} {\bibfnamefont {E.~M.}\ \bibnamefont
  {Rains}},\ }\href {\doibase 10.1023/A:1018615306992} {\bibfield  {journal}
  {\bibinfo  {journal} {Journal of Statistical Physics}\ }\textbf {\bibinfo
  {volume} {100}},\ \bibinfo {pages} {523} (\bibinfo {year}
  {2000})}\BibitemShut {NoStop}%
\bibitem [{\citenamefont {Kriecherbauer}\ and\ \citenamefont
  {Krug}(2010)}]{Kriecherbauer2010}%
  \BibitemOpen
  \bibfield  {author} {\bibinfo {author} {\bibfnamefont {T.}~\bibnamefont
  {Kriecherbauer}}\ and\ \bibinfo {author} {\bibfnamefont {J.}~\bibnamefont
  {Krug}},\ }\href {\doibase 10.1088/1751-8113/43/40/403001} {\bibfield
  {journal} {\bibinfo  {journal} {Journal of Physics A: Mathematical and
  Theoretical}\ }\textbf {\bibinfo {volume} {43}},\ \bibinfo {pages} {403001}
  (\bibinfo {year} {2010})}\BibitemShut {NoStop}%
\bibitem [{\citenamefont {Majumdar}(2006)}]{bouchaud2011complex}%
  \BibitemOpen
  \bibfield  {author} {\bibinfo {author} {\bibfnamefont {S.~N.}\ \bibnamefont
  {Majumdar}},\ }in\ \href {https://arxiv.org/abs/cond-mat/0701193} {\emph
  {\bibinfo {booktitle} {Complex Systems: Lecture Notes of the Les Houches
  Summer School 2006}}},\ \bibinfo {series and number} {Les Houches},\ \bibinfo
  {editor} {edited by\ \bibinfo {editor} {\bibfnamefont {J.}~\bibnamefont
  {Bouchaud}}, \bibinfo {editor} {\bibfnamefont {M.}~\bibnamefont
  {M{\'e}zard}}, \ and\ \bibinfo {editor} {\bibfnamefont {J.}~\bibnamefont
  {Dalibard}}}\ (\bibinfo  {publisher} {Elsevier Science},\ \bibinfo {year}
  {2006})\ Chap.~\bibinfo {chapter} {4}\BibitemShut {NoStop}%
\bibitem [{\citenamefont {Corvin}(2012)}]{corvin2012}%
  \BibitemOpen
  \bibfield  {author} {\bibinfo {author} {\bibfnamefont {I.}~\bibnamefont
  {Corvin}},\ }\href {\doibase 10.1142/S2010326311300014} {\bibfield  {journal}
  {\bibinfo  {journal} {Random Matrices: Theory and Applications}\ }\textbf
  {\bibinfo {volume} {01}},\ \bibinfo {pages} {1130001} (\bibinfo {year}
  {2012})}\BibitemShut {NoStop}%
\bibitem [{\citenamefont {Angel}\ \emph {et~al.}(2017)\citenamefont {Angel},
  \citenamefont {Balka},\ and\ \citenamefont {Peres}}]{angel2017increasing}%
  \BibitemOpen
  \bibfield  {author} {\bibinfo {author} {\bibfnamefont {O.}~\bibnamefont
  {Angel}}, \bibinfo {author} {\bibfnamefont {R.}~\bibnamefont {Balka}}, \ and\
  \bibinfo {author} {\bibfnamefont {Y.}~\bibnamefont {Peres}},\ }\href
  {\doibase 10.1017/S0305004116000797} {\bibfield  {journal} {\bibinfo
  {journal} {Mathematical Proceedings of the Cambridge Philosophical Society}\
  }\textbf {\bibinfo {volume} {163}},\ \bibinfo {pages} {173} (\bibinfo {year}
  {2017})}\BibitemShut {NoStop}%
\bibitem [{\citenamefont {Mendon{\c{c}}a}(2017)}]{mendoncca2017empirical}%
  \BibitemOpen
  \bibfield  {author} {\bibinfo {author} {\bibfnamefont {J.~R.~G.}\
  \bibnamefont {Mendon{\c{c}}a}},\ }\href@noop {} {\bibfield  {journal}
  {\bibinfo  {journal} {Journal of Physics A: Mathematical and Theoretical}\
  }\textbf {\bibinfo {volume} {50}},\ \bibinfo {pages} {08LT02} (\bibinfo
  {year} {2017})}\BibitemShut {NoStop}%
\bibitem [{\citenamefont {B\"orjes}\ \emph {et~al.}(2019)\citenamefont
  {B\"orjes}, \citenamefont {Schawe},\ and\ \citenamefont
  {Hartmann}}]{borjes2019large}%
  \BibitemOpen
  \bibfield  {author} {\bibinfo {author} {\bibfnamefont {J.}~\bibnamefont
  {B\"orjes}}, \bibinfo {author} {\bibfnamefont {H.}~\bibnamefont {Schawe}}, \
  and\ \bibinfo {author} {\bibfnamefont {A.~K.}\ \bibnamefont {Hartmann}},\
  }\href {\doibase 10.1103/PhysRevE.99.042104} {\bibfield  {journal} {\bibinfo
  {journal} {Phys. Rev. E}\ }\textbf {\bibinfo {volume} {99}},\ \bibinfo
  {pages} {042104} (\bibinfo {year} {2019})}\BibitemShut {NoStop}%
\bibitem [{\citenamefont {Mendon\ifmmode~\mbox{\c{c}}\else \c{c}\fi{}a}\ \emph
  {et~al.}(2020)\citenamefont {Mendon\ifmmode~\mbox{\c{c}}\else \c{c}\fi{}a},
  \citenamefont {Schawe},\ and\ \citenamefont
  {Hartmann}}]{mendonca2019asymptotic}%
  \BibitemOpen
  \bibfield  {author} {\bibinfo {author} {\bibfnamefont {J.~R.~G.}\
  \bibnamefont {Mendon\ifmmode~\mbox{\c{c}}\else \c{c}\fi{}a}}, \bibinfo
  {author} {\bibfnamefont {H.}~\bibnamefont {Schawe}}, \ and\ \bibinfo {author}
  {\bibfnamefont {A.~K.}\ \bibnamefont {Hartmann}},\ }\href {\doibase
  10.1103/PhysRevE.101.032102} {\bibfield  {journal} {\bibinfo  {journal}
  {Phys. Rev. E}\ }\textbf {\bibinfo {volume} {101}},\ \bibinfo {pages}
  {032102} (\bibinfo {year} {2020})}\BibitemShut {NoStop}%
\bibitem [{\citenamefont {Gopalan}\ \emph {et~al.}(2007)\citenamefont
  {Gopalan}, \citenamefont {Jayram}, \citenamefont {Krauthgamer},\ and\
  \citenamefont {Kumar}}]{gopalan2007}%
  \BibitemOpen
  \bibfield  {author} {\bibinfo {author} {\bibfnamefont {P.}~\bibnamefont
  {Gopalan}}, \bibinfo {author} {\bibfnamefont {T.~S.}\ \bibnamefont {Jayram}},
  \bibinfo {author} {\bibfnamefont {R.}~\bibnamefont {Krauthgamer}}, \ and\
  \bibinfo {author} {\bibfnamefont {R.}~\bibnamefont {Kumar}},\ }in\ \href@noop
  {} {\emph {\bibinfo {booktitle} {Proceedings of the Eighteenth Annual
  ACM-SIAM Symposium on Discrete Algorithms}}},\ \bibinfo {series and number}
  {SODA ’07}\ (\bibinfo  {publisher} {Society for Industrial and Applied
  Mathematics},\ \bibinfo {address} {USA},\ \bibinfo {year} {2007})\ p.\
  \bibinfo {pages} {318–327}\BibitemShut {NoStop}%
\bibitem [{\citenamefont {Bonomi}\ and\ \citenamefont
  {Xiong}(2016)}]{bonomi2016differentially}%
  \BibitemOpen
  \bibfield  {author} {\bibinfo {author} {\bibfnamefont {L.}~\bibnamefont
  {Bonomi}}\ and\ \bibinfo {author} {\bibfnamefont {L.}~\bibnamefont {Xiong}},\
  }\href@noop {} {\bibfield  {journal} {\bibinfo  {journal} {Transactions on
  Data Privacy}\ }\textbf {\bibinfo {volume} {9}},\ \bibinfo {pages} {73}
  (\bibinfo {year} {2016})}\BibitemShut {NoStop}%
\bibitem [{\citenamefont {Zhang}(2003)}]{Zhang2003}%
  \BibitemOpen
  \bibfield  {author} {\bibinfo {author} {\bibfnamefont {H.}~\bibnamefont
  {Zhang}},\ }\href {\doibase 10.1093/bioinformatics/btg168} {\bibfield
  {journal} {\bibinfo  {journal} {Bioinformatics}\ }\textbf {\bibinfo {volume}
  {19}},\ \bibinfo {pages} {1391} (\bibinfo {year} {2003})}\BibitemShut
  {NoStop}%
\bibitem [{\citenamefont {Hartmann}(2015)}]{practical_guide2015}%
  \BibitemOpen
  \bibfield  {author} {\bibinfo {author} {\bibfnamefont {A.~K.}\ \bibnamefont
  {Hartmann}},\ }\href@noop {} {\emph {\bibinfo {title} {{Big Practical Guide
  to Computer Simulations}}}}\ (\bibinfo  {publisher} {World Scientific},\
  \bibinfo {address} {Singapore},\ \bibinfo {year} {2015})\BibitemShut
  {NoStop}%
\bibitem [{\citenamefont {Hammersley}(1972)}]{hammersley1972}%
  \BibitemOpen
  \bibfield  {author} {\bibinfo {author} {\bibfnamefont {J.~M.}\ \bibnamefont
  {Hammersley}},\ }in\ \href
  {https://projecteuclid.org/euclid.bsmsp/1200514101} {\emph {\bibinfo
  {booktitle} {Proceedings of the Sixth Berkeley Symposium on Mathematical
  Statistics and Probability, Volume 1: Theory of Statistics}}}\ (\bibinfo
  {publisher} {University of California Press},\ \bibinfo {address} {Berkeley,
  Calif.},\ \bibinfo {year} {1972})\ pp.\ \bibinfo {pages}
  {345--394}\BibitemShut {NoStop}%
\bibitem [{\citenamefont {Houdré}\ and\ \citenamefont
  {Litherland}(2009)}]{houdre2009}%
  \BibitemOpen
  \bibfield  {author} {\bibinfo {author} {\bibfnamefont {C.}~\bibnamefont
  {Houdré}}\ and\ \bibinfo {author} {\bibfnamefont {T.~J.}\ \bibnamefont
  {Litherland}},\ }\enquote {\bibinfo {title} {On the longest increasing
  subsequence for finite and countable alphabets},}\ in\ \href {\doibase
  10.1214/09-IMSCOLL513} {\emph {\bibinfo {booktitle} {High Dimensional
  Probability V: The Luminy Volume}}},\ \bibinfo {series} {Collections}, Vol.\
  \bibinfo {volume} {Volume 5},\ \bibinfo {editor} {edited by\ \bibinfo
  {editor} {\bibfnamefont {C.}~\bibnamefont {Houdré}}, \bibinfo {editor}
  {\bibfnamefont {V.}~\bibnamefont {Koltchinskii}}, \bibinfo {editor}
  {\bibfnamefont {D.~M.}\ \bibnamefont {Mason}}, \ and\ \bibinfo {editor}
  {\bibfnamefont {M.}~\bibnamefont {Peligrad}}}\ (\bibinfo  {publisher}
  {Institute of Mathematical Statistics},\ \bibinfo {address} {Beachwood, Ohio,
  USA},\ \bibinfo {year} {2009})\ pp.\ \bibinfo {pages} {185--212}\BibitemShut
  {NoStop}%
\bibitem [{\citenamefont {Mallows}(1963)}]{Mallows1963problem}%
  \BibitemOpen
  \bibfield  {author} {\bibinfo {author} {\bibfnamefont {C.~L.}\ \bibnamefont
  {Mallows}},\ }\href@noop {} {\bibfield  {journal} {\bibinfo  {journal} {SIAM
  Review}\ }\textbf {\bibinfo {volume} {5}},\ \bibinfo {pages} {375} (\bibinfo
  {year} {1963})}\BibitemShut {NoStop}%
\bibitem [{\citenamefont {Chandramouli}\ and\ \citenamefont
  {Goldstein}(2014)}]{Chandramouli2014patience}%
  \BibitemOpen
  \bibfield  {author} {\bibinfo {author} {\bibfnamefont {B.}~\bibnamefont
  {Chandramouli}}\ and\ \bibinfo {author} {\bibfnamefont {J.}~\bibnamefont
  {Goldstein}},\ }in\ \href
  {https://www.microsoft.com/en-us/research/publication/patience-is-a-virtue-revisiting-merge-and-sort-on-modern-processors/}
  {\emph {\bibinfo {booktitle} {ACM SIGMOD International Conference on
  Management of Data (SIGMOD 2014)}}}\ (\bibinfo  {publisher} {ACM SIGMOD},\
  \bibinfo {year} {2014})\BibitemShut {NoStop}%
\bibitem [{\citenamefont {Bespamyatnikh}\ and\ \citenamefont
  {Segal}(2000)}]{Bespamyatnikh2007}%
  \BibitemOpen
  \bibfield  {author} {\bibinfo {author} {\bibfnamefont {S.}~\bibnamefont
  {Bespamyatnikh}}\ and\ \bibinfo {author} {\bibfnamefont {M.}~\bibnamefont
  {Segal}},\ }\href {\doibase 10.1016/S0020-0190(00)00124-1} {\bibfield
  {journal} {\bibinfo  {journal} {Information Processing Letters}\ }\textbf
  {\bibinfo {volume} {76}},\ \bibinfo {pages} {7 } (\bibinfo {year}
  {2000})}\BibitemShut {NoStop}%
\bibitem [{\citenamefont {Crochemore}\ and\ \citenamefont
  {Porat}(2010)}]{Crochemore2010}%
  \BibitemOpen
  \bibfield  {author} {\bibinfo {author} {\bibfnamefont {M.}~\bibnamefont
  {Crochemore}}\ and\ \bibinfo {author} {\bibfnamefont {E.}~\bibnamefont
  {Porat}},\ }\href {\doibase 10.1016/j.ic.2010.04.003} {\bibfield  {journal}
  {\bibinfo  {journal} {Information and Computation}\ }\textbf {\bibinfo
  {volume} {208}},\ \bibinfo {pages} {1054 } (\bibinfo {year}
  {2010})}\BibitemShut {NoStop}%
\bibitem [{\citenamefont {Arlotto}\ \emph {et~al.}(2015)\citenamefont
  {Arlotto}, \citenamefont {Nguyen},\ and\ \citenamefont
  {Steele}}]{Arlotto2015}%
  \BibitemOpen
  \bibfield  {author} {\bibinfo {author} {\bibfnamefont {A.}~\bibnamefont
  {Arlotto}}, \bibinfo {author} {\bibfnamefont {V.~V.}\ \bibnamefont {Nguyen}},
  \ and\ \bibinfo {author} {\bibfnamefont {J.~M.}\ \bibnamefont {Steele}},\
  }\href {\doibase 10.1016/j.spa.2015.03.009} {\bibfield  {journal} {\bibinfo
  {journal} {Stochastic Processes and their Applications}\ }\textbf {\bibinfo
  {volume} {125}},\ \bibinfo {pages} {3596 } (\bibinfo {year}
  {2015})}\BibitemShut {NoStop}%
\bibitem [{\citenamefont {Albert}\ \emph {et~al.}(2004)\citenamefont {Albert},
  \citenamefont {Golynski}, \citenamefont {Hamel}, \citenamefont
  {López-Ortiz}, \citenamefont {Rao},\ and\ \citenamefont
  {Safari}}]{Albert2004}%
  \BibitemOpen
  \bibfield  {author} {\bibinfo {author} {\bibfnamefont {M.~H.}\ \bibnamefont
  {Albert}}, \bibinfo {author} {\bibfnamefont {A.}~\bibnamefont {Golynski}},
  \bibinfo {author} {\bibfnamefont {A.~M.}\ \bibnamefont {Hamel}}, \bibinfo
  {author} {\bibfnamefont {A.}~\bibnamefont {López-Ortiz}}, \bibinfo {author}
  {\bibfnamefont {S.}~\bibnamefont {Rao}}, \ and\ \bibinfo {author}
  {\bibfnamefont {M.~A.}\ \bibnamefont {Safari}},\ }\href {\doibase
  10.1016/j.tcs.2004.03.057} {\bibfield  {journal} {\bibinfo  {journal}
  {Theoretical Computer Science}\ }\textbf {\bibinfo {volume} {321}},\ \bibinfo
  {pages} {405 } (\bibinfo {year} {2004})}\BibitemShut {NoStop}%
\bibitem [{\citenamefont {Li}\ \emph {et~al.}(2016)\citenamefont {Li},
  \citenamefont {Zou}, \citenamefont {Zhang},\ and\ \citenamefont
  {Zhao}}]{Youhuan2016}%
  \BibitemOpen
  \bibfield  {author} {\bibinfo {author} {\bibfnamefont {Y.}~\bibnamefont
  {Li}}, \bibinfo {author} {\bibfnamefont {L.}~\bibnamefont {Zou}}, \bibinfo
  {author} {\bibfnamefont {H.}~\bibnamefont {Zhang}}, \ and\ \bibinfo {author}
  {\bibfnamefont {D.}~\bibnamefont {Zhao}},\ }\href {\doibase
  10.14778/3021924.3021934} {\bibfield  {journal} {\bibinfo  {journal} {Proc.
  VLDB Endow.}\ }\textbf {\bibinfo {volume} {10}},\ \bibinfo {pages}
  {181–192} (\bibinfo {year} {2016})}\BibitemShut {NoStop}%
\bibitem [{\citenamefont {Salauyou}(2014)}]{stackoverflow}%
  \BibitemOpen
  \bibfield  {author} {\bibinfo {author} {\bibfnamefont {A.}~\bibnamefont
  {Salauyou}},\ }\href@noop {} {} (\bibinfo {year} {2014}),\ \bibinfo {note}
  {\url{https://stackoverflow.com/questions/22923646/number-of-all-longest-increasing-subsequences/22945390#22945390}}\BibitemShut
  {NoStop}%
\bibitem [{Note1()}]{Note1}%
  \BibitemOpen
  \bibinfo {note} {Other numbers is the same stack appear earlier in the same
  sequence but are larger, or they appear later but are smaller.}\BibitemShut
  {Stop}%
\bibitem [{\citenamefont {Hartmann}(2002)}]{Hartmann2002Sampling}%
  \BibitemOpen
  \bibfield  {author} {\bibinfo {author} {\bibfnamefont {A.~K.}\ \bibnamefont
  {Hartmann}},\ }\href {\doibase 10.1103/PhysRevE.65.056102} {\bibfield
  {journal} {\bibinfo  {journal} {Phys. Rev. E}\ }\textbf {\bibinfo {volume}
  {65}},\ \bibinfo {pages} {056102} (\bibinfo {year} {2002})}\BibitemShut
  {NoStop}%
\bibitem [{\citenamefont {Hartmann}(2011)}]{Hartmann2011}%
  \BibitemOpen
  \bibfield  {author} {\bibinfo {author} {\bibfnamefont {A.~K.}\ \bibnamefont
  {Hartmann}},\ }\href {\doibase 10.1140/epjb/e2011-10836-4} {\bibfield
  {journal} {\bibinfo  {journal} {The European Physical Journal B}\ }\textbf
  {\bibinfo {volume} {84}},\ \bibinfo {pages} {627} (\bibinfo {year}
  {2011})}\BibitemShut {NoStop}%
\bibitem [{\citenamefont {Schawe}\ and\ \citenamefont
  {Hartmann}(2019)}]{schawe2019large}%
  \BibitemOpen
  \bibfield  {author} {\bibinfo {author} {\bibfnamefont {H.}~\bibnamefont
  {Schawe}}\ and\ \bibinfo {author} {\bibfnamefont {A.~K.}\ \bibnamefont
  {Hartmann}},\ }\href {\doibase 10.1140/epjb/e2019-90667-y} {\bibfield
  {journal} {\bibinfo  {journal} {The European Physical Journal B}\ }\textbf
  {\bibinfo {volume} {92}},\ \bibinfo {pages} {73} (\bibinfo {year}
  {2019})}\BibitemShut {NoStop}%
\bibitem [{\citenamefont {Schawe}\ \emph {et~al.}(2018)\citenamefont {Schawe},
  \citenamefont {Hartmann},\ and\ \citenamefont
  {Majumdar}}]{schawe2018avoiding}%
  \BibitemOpen
  \bibfield  {author} {\bibinfo {author} {\bibfnamefont {H.}~\bibnamefont
  {Schawe}}, \bibinfo {author} {\bibfnamefont {A.~K.}\ \bibnamefont
  {Hartmann}}, \ and\ \bibinfo {author} {\bibfnamefont {S.~N.}\ \bibnamefont
  {Majumdar}},\ }\href {\doibase 10.1103/PhysRevE.97.062159} {\bibfield
  {journal} {\bibinfo  {journal} {Phys. Rev. E}\ }\textbf {\bibinfo {volume}
  {97}},\ \bibinfo {pages} {062159} (\bibinfo {year} {2018})}\BibitemShut
  {NoStop}%
\bibitem [{\citenamefont {Hartmann}(2014)}]{Hartmann2014high}%
  \BibitemOpen
  \bibfield  {author} {\bibinfo {author} {\bibfnamefont {A.~K.}\ \bibnamefont
  {Hartmann}},\ }\href {\doibase 10.1103/PhysRevE.89.052103} {\bibfield
  {journal} {\bibinfo  {journal} {Phys. Rev. E}\ }\textbf {\bibinfo {volume}
  {89}},\ \bibinfo {pages} {052103} (\bibinfo {year} {2014})}\BibitemShut
  {NoStop}%
\bibitem [{\citenamefont {Hartmann}\ \emph {et~al.}(2018)\citenamefont
  {Hartmann}, \citenamefont {Doussal}, \citenamefont {Majumdar}, \citenamefont
  {Rosso},\ and\ \citenamefont {Schehr}}]{kpz2018}%
  \BibitemOpen
  \bibfield  {author} {\bibinfo {author} {\bibfnamefont {A.~K.}\ \bibnamefont
  {Hartmann}}, \bibinfo {author} {\bibfnamefont {P.~L.}\ \bibnamefont
  {Doussal}}, \bibinfo {author} {\bibfnamefont {S.~N.}\ \bibnamefont
  {Majumdar}}, \bibinfo {author} {\bibfnamefont {A.}~\bibnamefont {Rosso}}, \
  and\ \bibinfo {author} {\bibfnamefont {G.}~\bibnamefont {Schehr}},\
  }\href@noop {} {\bibfield  {journal} {\bibinfo  {journal} {Europhys. Lett.}\
  }\textbf {\bibinfo {volume} {121}},\ \bibinfo {pages} {67004} (\bibinfo
  {year} {2018})}\BibitemShut {NoStop}%
\bibitem [{\citenamefont {Metropolis}\ \emph {et~al.}(1953)\citenamefont
  {Metropolis}, \citenamefont {Rosenbluth}, \citenamefont {Rosenbluth},
  \citenamefont {Teller},\ and\ \citenamefont
  {Teller}}]{metropolis1953equation}%
  \BibitemOpen
  \bibfield  {author} {\bibinfo {author} {\bibfnamefont {N.}~\bibnamefont
  {Metropolis}}, \bibinfo {author} {\bibfnamefont {A.~W.}\ \bibnamefont
  {Rosenbluth}}, \bibinfo {author} {\bibfnamefont {M.~N.}\ \bibnamefont
  {Rosenbluth}}, \bibinfo {author} {\bibfnamefont {A.~H.}\ \bibnamefont
  {Teller}}, \ and\ \bibinfo {author} {\bibfnamefont {E.}~\bibnamefont
  {Teller}},\ }\href@noop {} {\bibfield  {journal} {\bibinfo  {journal} {The
  journal of chemical physics}\ }\textbf {\bibinfo {volume} {21}},\ \bibinfo
  {pages} {1087} (\bibinfo {year} {1953})}\BibitemShut {NoStop}%
\bibitem [{\citenamefont {D'Agostino}\ and\ \citenamefont
  {Pearson}(1973)}]{Agostino1973}%
  \BibitemOpen
  \bibfield  {author} {\bibinfo {author} {\bibfnamefont {R.}~\bibnamefont
  {D'Agostino}}\ and\ \bibinfo {author} {\bibfnamefont {E.~S.}\ \bibnamefont
  {Pearson}},\ }\href@noop {} {\bibfield  {journal} {\bibinfo  {journal}
  {Biometrika}\ }\textbf {\bibinfo {volume} {60}},\ \bibinfo {pages} {613}
  (\bibinfo {year} {1973})}\BibitemShut {NoStop}%
\bibitem [{\citenamefont {Press}\ \emph {et~al.}(2007)\citenamefont {Press},
  \citenamefont {Teukolsky}, \citenamefont {Vetterling},\ and\ \citenamefont
  {Flannery}}]{press2007numerical}%
  \BibitemOpen
  \bibfield  {author} {\bibinfo {author} {\bibfnamefont {W.~H.}\ \bibnamefont
  {Press}}, \bibinfo {author} {\bibfnamefont {S.~A.}\ \bibnamefont
  {Teukolsky}}, \bibinfo {author} {\bibfnamefont {W.~T.}\ \bibnamefont
  {Vetterling}}, \ and\ \bibinfo {author} {\bibfnamefont {B.~P.}\ \bibnamefont
  {Flannery}},\ }\href@noop {} {\emph {\bibinfo {title} {Numerical recipes 3rd
  edition: The art of scientific computing}}}\ (\bibinfo  {publisher}
  {Cambridge university press},\ \bibinfo {year} {2007})\BibitemShut {NoStop}%
\bibitem [{\citenamefont {{Virtanen}}\ \emph {et~al.}(2019)\citenamefont
  {{Virtanen}}, \citenamefont {{Gommers}}, \citenamefont {{Oliphant}},
  \citenamefont {{Haberland}}, \citenamefont {{Reddy}}, \citenamefont
  {{Cournapeau}}, \citenamefont {{Burovski}}, \citenamefont {{Peterson}},
  \citenamefont {{Weckesser}}, \citenamefont {{Bright}}, \citenamefont {{van
  der Walt}}, \citenamefont {{Brett}}, \citenamefont {{Wilson}}, \citenamefont
  {{Jarrod Millman}}, \citenamefont {{Mayorov}}, \citenamefont {{Nelson}},
  \citenamefont {{Jones}}, \citenamefont {{Kern}}, \citenamefont {{Larson}},
  \citenamefont {{Carey}}, \citenamefont {{Polat}}, \citenamefont {{Feng}},
  \citenamefont {{Moore}}, \citenamefont {{Vand erPlas}}, \citenamefont
  {{Laxalde}}, \citenamefont {{Perktold}}, \citenamefont {{Cimrman}},
  \citenamefont {{Henriksen}}, \citenamefont {{Quintero}}, \citenamefont
  {{Harris}}, \citenamefont {{Archibald}}, \citenamefont {{Ribeiro}},
  \citenamefont {{Pedregosa}}, \citenamefont {{van Mulbregt}},\ and\
  \citenamefont {{Contributors}}}]{scipy}%
  \BibitemOpen
  \bibfield  {author} {\bibinfo {author} {\bibfnamefont {P.}~\bibnamefont
  {{Virtanen}}}, \bibinfo {author} {\bibfnamefont {R.}~\bibnamefont
  {{Gommers}}}, \bibinfo {author} {\bibfnamefont {T.~E.}\ \bibnamefont
  {{Oliphant}}}, \bibinfo {author} {\bibfnamefont {M.}~\bibnamefont
  {{Haberland}}}, \bibinfo {author} {\bibfnamefont {T.}~\bibnamefont
  {{Reddy}}}, \bibinfo {author} {\bibfnamefont {D.}~\bibnamefont
  {{Cournapeau}}}, \bibinfo {author} {\bibfnamefont {E.}~\bibnamefont
  {{Burovski}}}, \bibinfo {author} {\bibfnamefont {P.}~\bibnamefont
  {{Peterson}}}, \bibinfo {author} {\bibfnamefont {W.}~\bibnamefont
  {{Weckesser}}}, \bibinfo {author} {\bibfnamefont {J.}~\bibnamefont
  {{Bright}}}, \bibinfo {author} {\bibfnamefont {S.~J.}\ \bibnamefont {{van der
  Walt}}}, \bibinfo {author} {\bibfnamefont {M.}~\bibnamefont {{Brett}}},
  \bibinfo {author} {\bibfnamefont {J.}~\bibnamefont {{Wilson}}}, \bibinfo
  {author} {\bibfnamefont {K.}~\bibnamefont {{Jarrod Millman}}}, \bibinfo
  {author} {\bibfnamefont {N.}~\bibnamefont {{Mayorov}}}, \bibinfo {author}
  {\bibfnamefont {A.~R.~J.}\ \bibnamefont {{Nelson}}}, \bibinfo {author}
  {\bibfnamefont {E.}~\bibnamefont {{Jones}}}, \bibinfo {author} {\bibfnamefont
  {R.}~\bibnamefont {{Kern}}}, \bibinfo {author} {\bibfnamefont
  {E.}~\bibnamefont {{Larson}}}, \bibinfo {author} {\bibfnamefont
  {C.}~\bibnamefont {{Carey}}}, \bibinfo {author} {\bibfnamefont
  {{\.I}.}~\bibnamefont {{Polat}}}, \bibinfo {author} {\bibfnamefont
  {Y.}~\bibnamefont {{Feng}}}, \bibinfo {author} {\bibfnamefont {E.~W.}\
  \bibnamefont {{Moore}}}, \bibinfo {author} {\bibfnamefont {J.}~\bibnamefont
  {{Vand erPlas}}}, \bibinfo {author} {\bibfnamefont {D.}~\bibnamefont
  {{Laxalde}}}, \bibinfo {author} {\bibfnamefont {J.}~\bibnamefont
  {{Perktold}}}, \bibinfo {author} {\bibfnamefont {R.}~\bibnamefont
  {{Cimrman}}}, \bibinfo {author} {\bibfnamefont {I.}~\bibnamefont
  {{Henriksen}}}, \bibinfo {author} {\bibfnamefont {E.~A.}\ \bibnamefont
  {{Quintero}}}, \bibinfo {author} {\bibfnamefont {C.~R.}\ \bibnamefont
  {{Harris}}}, \bibinfo {author} {\bibfnamefont {A.~M.}\ \bibnamefont
  {{Archibald}}}, \bibinfo {author} {\bibfnamefont {A.~H.}\ \bibnamefont
  {{Ribeiro}}}, \bibinfo {author} {\bibfnamefont {F.}~\bibnamefont
  {{Pedregosa}}}, \bibinfo {author} {\bibfnamefont {P.}~\bibnamefont {{van
  Mulbregt}}}, \ and\ \bibinfo {author} {\bibfnamefont {S.~.~.}\ \bibnamefont
  {{Contributors}}},\ }\href@noop {} {\bibfield  {journal} {\bibinfo  {journal}
  {arXiv e-prints}\ ,\ \bibinfo {eid} {arXiv:1907.10121}} (\bibinfo {year}
  {2019})},\ \Eprint {http://arxiv.org/abs/1907.10121} {arXiv:1907.10121
  [cs.MS]} \BibitemShut {NoStop}%
\bibitem [{\citenamefont {Schawe}\ and\ \citenamefont
  {Hartmann}(2020)}]{schawe2020large}%
  \BibitemOpen
  \bibfield  {author} {\bibinfo {author} {\bibfnamefont {H.}~\bibnamefont
  {Schawe}}\ and\ \bibinfo {author} {\bibfnamefont {A.~K.}\ \bibnamefont
  {Hartmann}},\ }\href@noop {} {\bibfield  {journal} {\bibinfo  {journal}
  {arXiv preprint arXiv:2003.03415}\ } (\bibinfo {year} {2020})}\BibitemShut
  {NoStop}%
\bibitem [{\citenamefont {Touchette}(2009)}]{Touchette2009large}%
  \BibitemOpen
  \bibfield  {author} {\bibinfo {author} {\bibfnamefont {H.}~\bibnamefont
  {Touchette}},\ }\href {\doibase 10.1016/j.physrep.2009.05.002} {\bibfield
  {journal} {\bibinfo  {journal} {Physics Reports}\ }\textbf {\bibinfo {volume}
  {478}},\ \bibinfo {pages} {1 } (\bibinfo {year} {2009})}\BibitemShut
  {NoStop}%
\end{thebibliography}%

\end{document}